\documentclass[10pt, showpacs,preprintnumbers, pra]{revtex4}
\usepackage{graphicx}
\usepackage{floatrow}
\usepackage[caption=false]{subfig}
\usepackage{xcolor}
\usepackage{amsmath}
\usepackage{amssymb}
\usepackage{mleftright}
\usepackage{dsfont}
\usepackage{floatrow}
\mleftright
\def\bc{\begin{center}}
\def\ec{\end{center}}

\def\beq{\begin{equation}}
\def\eeq{\end{equation}}

\begin{document}

\title{{Can the dynamical Lamb effect be observed in a superconducting circuit?}}
\author{Mirko Amico$^{1,2}$, Oleg L. Berman$^{1,2}$ and Roman Ya. Kezerashvili$^{1,2}$}
\affiliation{\mbox{$^{1}$Physics Department, New York City College
of Technology, The City University of New York,} \\
Brooklyn, NY 11201, USA \\
\mbox{$^{2}$The Graduate School and University Center, The
City University of New York,} \\
New York, NY 10016, USA}

\begin{abstract}

The dynamical Lamb effect is predicted to arise in superconducting circuits when the coupling of a superconducting qubit with a resonator is periodically switched "on" and "off" nonadiabatically. We show that by using a superconducting circuit which allows to switch between longitudinal and transverse coupling of a qubit to a resonator, it is possible of to observe the dynamical Lamb effect. {The switching between longitudinal and transverse coupling can be achieved by modulating the magnetic flux through the circuit loops.} By solving the Schr\"{o}dinger equation for a qubit coupled to a resonator, we calculate the time evolution of the probability of excitation of the qubit and the creation of $n$ photons in the resonator due to the dynamical Lamb effect. The probability is maximum when the coupling is periodically switched between longitudinal and transverse using a square-wave or sinusoidal modulation of the magnetic flux with frequency equal to the sum of the average qubit and photon transition frequencies.
\end{abstract}
\pacs{}


\maketitle

\section{Introduction}
\label{intro}

According to quantum field theory, the vacuum is filled with virtual particles which can be turned into real ones by {specific external perturbations \cite{fulling}}. Phenomena of this kind are commonly referred to as quantum vacuum phenomena. Several quantum vacuum phenomena related to the peculiar nature of the quantum vacuum have been predicted \cite{casimir, bethe}, some of which have been experimentally found \cite{lamb}. Other examples include the dynamical Casimir effect \cite{moore}, {that is} the creation of real photons from the vacuum due to the fast change in boundary conditions of a cavity, and the dynamical Lamb effect \cite{dle}, {which is} the excitation of an atom in a cavity, along with the creation of photons, due to the sudden change of its Lamb shift. To obtain an instantaneous change of the Lamb shift of the atom, the boundary conditions of the cavity must be changed nonadiabatically \cite{dle, berman, amico2017, amico2018}.

Recently, the dynamical Casimir effect has been experimentally observed in superconducting circuits \cite{wilson, lahte}. The latter provide a way to model atoms and cavities using Josephson junctions and superconducting transmission lines. The advantage of a superconducting circuit setup over real atoms and cavities lies in the possibility of tuning the parameters of the system in a short time interval, allowing us to enter the nonadiabatic regime where the mentioned quantum vacuum phenomena arise. 

As noted in Ref. \cite{shapiro}, the dynamical Lamb effect could be observed in a superconducting circuit as well. The nonadiabatic change in boundary conditions of the cavity needed for the dynamical Lamb effect to arise can be obtained by switching "on" and "off" the coupling of a qubit with a resonator. Furthermore, the periodic switching "on" and "off" of the qubit/resonator coupling leads to a dramatic increase in the probability of excitation of the qubit \cite{shapiro, zhukov, remizov, amico2018}. In fact, the dynamical aspects of the Lamb shift in the energy levels of tunable superconducting circuits have already been investigated both theoretically \cite{gramich1, gramich2} and experimentally \cite{silveri}. Here we focus on the case of the particular tuning required to enter the nonadiabatic regime which gives rise to the dynamical Lamb effect.

In Refs. \cite{richer2, richer}, it was shown that it is possible to design a superconducting circuit where the qubit/resonator coupling is switched between longitudinal and transverse by modulating the magnetic flux through the circuit loops. A qubit/resonator system longitudinally coupled can be seen as a decoupled system with renormalized energy levels \cite{lang}. Whereas in a qubit/resonator system with transverse coupling the qubit and the photons interact. Therefore, we suggest the possibility of observing the dynamical Lamb effect by adopting the circuit designed in Ref. \cite{richer} and periodically {switching} between longitudinal and transverse qubit/resonator coupling. This effectively {corresponds} to periodically {switching} "on" and "off" of the qubit/resonator coupling, which has been shown to give rise to the dynamical Lamb effect \cite{shapiro, zhukov, remizov, amico2018}.

To demonstrate the presence of the dynamical Lamb effect, we calculate the probability of excitation of the qubit and the probability of creation of photons in the resonator by solving the Schr\"{o}dinger equation. The calculations show that {the {probabilities} of excitation of the qubit and creation of photons due to the dynamical Lamb effect {reach their {maximum} values} when the coupling is periodically switched between transverse and longitudinal using a square-wave or sinusoidal modulation of the magnetic flux with frequency equal to the sum of the average qubit and photon transition frequencies.}

The article is organized as follows. In Sec. \ref{lg_tr} the Hamiltonian of a qubit/resonator system with longitudinal or transverse coupling is specified. In Sec. \ref{s_circuit}, a superconducting circuit which allows for the switching between a longitudinally coupled Hamiltonian and a transverse one is introduced. We show how to switch between longitudinal and transverse coupling through the modulation of the magnetic flux threading the circuit. The results of numerical calculations of the time evolution of the probability of excitation of the qubit and the photons for different modulation of the magnetic flux are given in Sec. \ref{res}. The conclusions follow in Sec. \ref{conc}.

\section{Longitudinal and transverse coupling}
\label{lg_tr}

As a first step, let us show how a system with longitudinal qubit/resonator
coupling can be seen as an uncoupled system, in contrast to the case of transverse qubit/resonator
coupling. The Hamiltonians of a qubit longitudinally {$\hat{{H}}_{L}$} and
transversely {$\hat{{H}}_{T}$} coupled to a resonator, respectively, can be written as 


\begin{equation}
\hat{{H}}_{L}=\hbar \omega _{0}\hat{\sigma}^{+}\hat{\sigma}^{-}+ \hbar\omega _{r}\hat{a}%
^{\dagger }\hat{a}+\hbar g_{zx}\hat{\sigma}_{z}\left( \hat{a}^{\dagger }+\hat{a}%
\right) ,  \label{longitudinal}
\end{equation}


\begin{equation}
\hat{{H}}_{T}=\hbar \omega _{0}\hat{\sigma}^{+}\hat{\sigma}^{-}+ \hbar \omega _{r}\hat{a}%
^{\dagger }\hat{a}+ \hbar g_{xx}\hat{\sigma}_{x}\left( \hat{a}^{\dagger }+\hat{a}%
\right) ,  \label{transverse}
\end{equation}

\noindent where $\omega _{0}$ is the transition frequency of the qubit, $%
\omega _{r}$ is the frequency of the photons in the resonator, {$\hat{\sigma}^{+} = \frac{\hat{\sigma}_x + i \hat{\sigma}_y }{2}$, $%
\hat{\sigma}^{-} = \frac{\hat{\sigma}_x - i \hat{\sigma}_y }{2}$ and $\hat{a}^{\dagger }$, $\hat{a}$ are the creation and
annihilation operators for excitations of qubit and photons,
respectively, $\hat{\sigma}_{x}$, $\hat{\sigma}_{y}$ and $\hat{\sigma}_{z}$ are the Pauli $x$, $y$
and $z$ operators}, while $g_{zx}$ and $g_{xx}$ are the longitudinal and transverse coupling
{strengths}, respectively. Applying an appropriate unitary transformation \cite{lang, billangeon}, the Hamiltonian (\ref{longitudinal}) can be written in a diagonal form as


\begin{equation}  \label{longitudinal2}
\hat{{H}}^{\prime}_{L} = \hbar \omega_0 \hat{\sigma}^{+}\hat{\sigma}^{-} + \hbar \omega_r
\hat{a}^{\dagger}\hat{a} - \frac{\hbar^2 g_{zx}^{2}}{\omega_r} \hat{I} ,
\end{equation}


\noindent
{where $\hat{I}$ is the identity operator.
{Since $\hat{{H}}^{\prime}_{L}$ and $\hat{{H}}_{L}$ are related by a unitary transformation, their eigenvalues are the same and they describe a qubit and a resonator with the same transition frequencies and they describe a qubit and a resonator with the same transition frequencies. {Therefore, the two Hamiltonians describe systems which are characterized by the same observables.}} However, in (\ref{longitudinal2}) the qubit is now decoupled from the resonator and the zero-point energy is renormalized. In this case, the Lamb shift of the qubit is absent. On the other hand, one can not diagonalize Hamiltonian (\ref{transverse}) by any unitary
transformation, and, therefore, the qubit/resonator
coupling cannot be eliminated. The latter implies that the energy levels of the qubit will be affected by the Lamb shift.
So, we can regard the system with longitudinal coupling given by Eq. (\ref{longitudinal}) as a system of a qubit and a resonator with the
qubit/resonator coupling turned "off" and the system with transverse coupling
defined by Hamiltonian (\ref{transverse}) as the same qubit and resonator with the qubit/resonator
coupling turned "on". Thus, the switching between these two coupling regimes involves a change in the Lamb shift of the qubit.}

\section{Superconducting circuit with tunable qubit/resonator coupling}
\label{s_circuit}

\begin{figure}[]
\includegraphics[height=6cm]{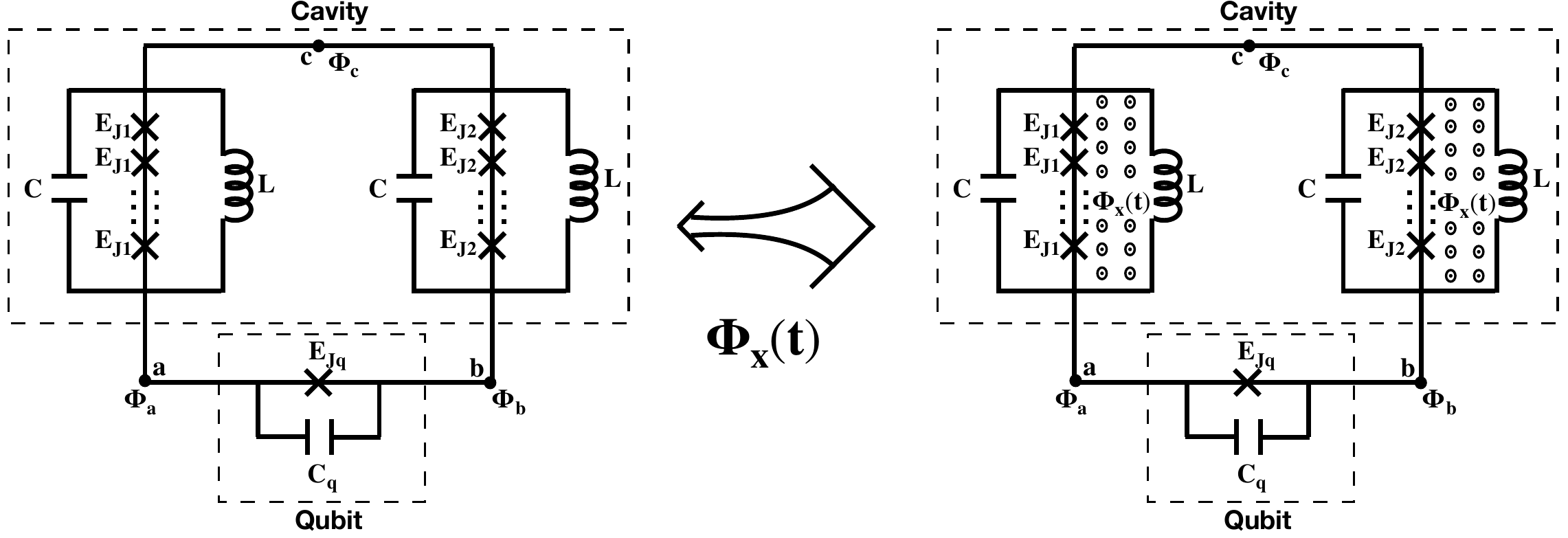}
\caption{Superconducting circuit for a qubit coupled to a resonator with
tunable qubit/resonator coupling. By turning "on" and "off" the magnetic flux $\Phi_x (t)$ we can switch between a description of the circuit in terms of a transversely coupled Hamiltonian and a longitudinal one.}
\label{circuit}
\end{figure}

Let us consider the circuit in Fig. \ref{circuit} and define the branch fluxes associated with the qubit and the resonator, as $\Phi_q = \Phi_a - \Phi_b$ and $\Phi_r = \Phi_a + \Phi_b - 2\Phi_c$, respectively, where $\Phi_a$, $\Phi_b$ and $\Phi_c$ are the magnetic fluxes at the nodes $a$, $b$ and $c$. Following Ref. \cite{devoret}, one can write the Lagrangian for the circuit {in Fig. \ref{circuit}} by adding the contributions of each element in terms of the branch fluxes {\cite{richer}}


\begin{equation}
\begin{split}
\label{flux}
\mathcal{L} =  \left( \frac{2C_q + C}{4} \dot{\Phi}^{2}_q + \frac{C}{2} \dot{\Phi}^{2}_r \right) - \frac{1}{4L} \left( \Phi_q^{2} + \Phi_r^{2} \right) + E_{J_{q}} \cos \left( \frac{2\pi}{\Phi_0} \Phi_q  \right) + \\
 + k E_{J1} \cos \left( \frac{2\pi}{\Phi_0} \left( \frac{\Phi_q + \Phi_r }{2k} + \frac{\Phi_x(t)}{k}\right) \right) + k E_{J2} \cos \left( \frac{2\pi}{\Phi_0} \left( \frac{\Phi_q - \Phi_r }{2k} + \frac{\Phi_x(t)}{k} \right) \right) .
\end{split}
\end{equation}


\noindent
In Eq. (\ref{flux}), $\Phi_x(t)$ is the external magnetic flux threading the areas enclosed by the left and right loops, $k$ is the number of Josephson junctions in a branch of the circuit, which the same in each branch, $C$ and $L$ are the capacitance and the inductance of the loops, respectively, $E_{J1}$ and  $E_{J2}$ are the Josephson energies of the junctions in each branch, $E_{Jq}$ the Josephson energy of the qubit junction and $C_q$ its capacitance. The Hamiltonian of the system can be found by taking the Legendre transform of the Lagrangian: $\mathcal{H} = \sum_{i=1}^{N} \frac{d \mathcal{L} }{d \dot{\Phi}_i }  \dot{\Phi}_i - \mathcal{L}$, where $i = q,r$ are the indices corresponding to the qubit and resonator flux variables, respectively. This leads to the following Hamiltonian for the circuit

\begin{equation}
\begin{split}
\label{ham}
\mathcal{H}(t) =   \frac{1}{2C_q + C} Q_q^{2} + \frac{1}{C} Q_r^{2}  + \frac{1}{4L} \left( \Phi_q^{2} + \Phi_r^{2} \right) - E_{J_{q}} \cos \left( \frac{2\pi}{\Phi_0} \Phi_q  \right) + \\
 - k E_{J1} \cos \left( \frac{2\pi}{\Phi_0} \left( \frac{\Phi_q + \Phi_r }{2k} + \frac{\Phi_x(t)}{k}\right) \right) - k E_{J2} \cos \left( \frac{2\pi}{\Phi_0} \left( \frac{\Phi_q - \Phi_r }{2k} + \frac{\Phi_x(t)}{k} \right) \right) .
\end{split}
\end{equation}

A quantum mechanical model of the circuit can be obtained from its classical Hamiltonian by applying the standard procedure of second quantization for the qubit and resonator variables separately \cite{richer}. 
For example, the quantum mechanical model for the resonator is obtained from Hamiltonian (\ref{ham}) by setting $Q_q=0$ and $\Phi_q = 0$, expanding the cosine terms up to second order in $\Phi_r$, and expressing the resonator's variables $Q_r$ and $\Phi_r$ in terms of the operators of creation and annihilation of photons in the resonator $\hat{a}^{\dagger}$, $\hat{a}$, respectively,

\begin{equation}
\label{quantization_res}
\Phi_r = \left( \hbar^2 \frac{L}{C\left(1+\eta\right)}  \right)^{\frac{1}{4}} \left( \hat{a} + \hat{a}^{\dagger}\right) ,  \; \; \; \; \; \; \; \; \; \; \; \; \; \; \; \; \; \; \; \; \; \; \; \; \; \; \; Q_r =  \left( \left(\frac{\hbar}{2}\right)^2 \frac{C\left(1+\eta\right)}{L}  \right)^{\frac{1}{4}} i \left( \hat{a}^{\dagger} - \hat{a}\right) ,
\end{equation}

\noindent
which give

\begin{equation}
\label{resonator_q_ham}
\hat{\mathcal{H}}_r = \hbar \omega_r \left( \hat{a}^{\dagger}\hat{a} + \frac{1}{2} \right) ,
\end{equation}

\noindent
where $\omega_r = \sqrt{\frac{1+\eta}{LC}}$ is the transition frequency between the energy levels of the system and $\eta$ is a dimensionless parameter, defined in Table \ref{tab1}, which accounts for the flux-dependence of the system. The Hamiltonian (\ref{resonator_q_ham}) is the Hamiltonian of a harmonic oscillator. The operators of creation and annihilation of photons in the resonator are bosonic operators which satisfy the commutation relation $\left[\hat{a}, \hat{a}^{\dagger} \right] = 1$. With the definitions given in Eq. (\ref{quantization_res}), and the commutation relation for $\hat{a}^{\dagger}$ and $\hat{a}$, one can prove that the variables $\Phi_r$ and $Q_r$ satisfy the commutation relation for conjugate variables $\left[\Phi_r, Q_r \right] = i \hbar$.
One can do the same for the qubit variables, starting from Hamiltonian (\ref{ham}), setting $Q_r=0$ and $\Phi_r = 0$, expanding the cosine terms up to second order in $\Phi_q$, and introducing the operators of creation and annihilation of resonator excitation in terms of $Q_q$ and $\Phi_q$, 

\begin{equation}
\label{quantization_qub}
\Phi_q = \left( \frac{\Phi_0}{2\pi}\right) \left( \frac{2e^2}{2C_q + C} \frac{1}{E_{Jq} + \left( \frac{\Phi_0}{2\pi} \right)^2 \frac{1+\eta}{2L} }  \right)^{\frac{1}{4}} \left( \hat{b} + \hat{b}^{\dagger}\right) ,  \; \; \; \; \; \; \; \; \; \; \; \; Q_q =  e \left( \left( E_{Jq} + \left( \frac{\Phi_0}{2\pi} \right)^2 \frac{1+\eta}{2L} \right) \frac{2C_q + C}{2e^2}  \right)^{\frac{1}{4}} i \left( \hat{b}^{\dagger} - \hat{b}\right) ,
\end{equation}

\noindent
which give the following quantum mechanical Hamiltonian

\begin{equation}
\label{qubit_q_ham}
\hat{\mathcal{H}}_q = \hbar \omega_q \left( \hat{b}^{\dagger}\hat{b} + \frac{1}{2} \right) ,
\end{equation}

\noindent
where $\omega_q = \frac{\sqrt{  8 \left(E_{Jq} + \left( \frac{\Phi_0}{2\pi} \right)^2 \frac{1+\eta}{2L}\right) \frac{2e^2}{2C_q + C}  }}{\hbar}$ is the transition frequency between the first two energy levels of the system. The operators of creation and annihilation of qubit excitation are also taken to be bosonic operators satisfying the commutation relation $\left[\hat{b}, \hat{b}^{\dagger} \right] = 1$. Again, one can prove that the variables $\Phi_q$ and $Q_q$ satisfy the commutation relation for conjugate variables $\left[\Phi_q, Q_q \right] = i \hbar$ by using the commutation relation for $\hat{b}^{\dagger}$ and $\hat{b}$, together with the definitions given in Eq. (\ref{quantization_qub}). Since we consider only two accessible levels, we replace the the creation and annihilation operators $\hat{b}$ and $\hat{b}^{\dagger}$ with $\hat{\sigma}^{+}$ and $\hat{\sigma}^{-} $, which are used to describe excitations in a two-level system. The transition frequency between the first two levels is also adjusted to take into account the anharmonicity by replacing $\omega_q$ with $\omega_0$.  Therefore, we rewrite the Hamiltonian (\ref{qubit_q_ham}) as

\begin{equation}
\label{qubit_q_ham2}
\hat{\mathcal{H}}'_q = \hbar \omega_0 \left( \hat{\sigma}^{+} \hat{\sigma}^{-}  + \frac{1}{2} \right) ,
\end{equation}

\noindent
Hamiltonian (\ref{qubit_q_ham2}) is the Hamiltonian of a quantum two-level system.
To obtain a quantum mechanical Hamiltonian of the system, one can substitute the expressions for the resonator and qubit variables given in Eqs. (\ref{quantization_res}) and (\ref{quantization_qub}), respectively, into Hamiltonian (\ref{ham}). In this way, one can also express the terms in Hamiltonian (\ref{ham}) which involve both resonator and qubit variables in the argument of the cosine, thus coupling those variables, in terms of creation and annihilation operators of the photons excited in the resonator and qubit's excitation. Thus, getting


\begin{equation}
\begin{split}
\label{full_H}
\hat{\mathcal{H}} (t) = \hbar \omega_r(t) \left( \hat{a}^{\dagger}\hat{a} + \frac{1}{2}\right) +
\hbar \frac{\omega_0(t)}{2} \hat{\sigma}_z + \hbar g_{xx}(t) \hat{\sigma}_{x} \left( \hat{a} 
^{\dagger} + \hat{a} \right) +\hbar  g_{zz}(t) \hat{\sigma}_{z} \left( \hat{a}%
^{\dagger} + \hat{a} \right)^2 + \\
 + \hbar g_{zx}(t) \hat{\sigma}_{z} \left( \hat{a}%
^{\dagger} + \hat{a} \right) +\hbar  g_{xz}(t) \hat{\sigma}_{x} \left( \hat{a}%
^{\dagger} + \hat{a} \right)^2 ,
\end{split}%
\end{equation}


\noindent
where $\omega_r (t)$ is the transition frequency of the resonator, $\omega_0 (t)$ is the transition frequency of the qubit and $g_{xx} (t)$, $g_{zz}(t)$, $g_{zx}(t)$ and $g_{xz}(t)$ are the coupling strengths. The expressions of each of the parameters in Hamiltonian (\ref{full_H}) are given in Table \ref{tab1} in the Appendix. It is important to note that all these parameters depend on time through their dependence on the external magnetic flux $\Phi _{x}(t)$.

\subsection{Square-wave modulation}
\label{sq}

\begin{table}[htb]


\caption{Instantaneous values of the parameters given in Table \ref{tab1} for the case of square-wave modulation of the external magnetic flux $\Phi _{x}$.}
\centering
\begin{tabular}{|l|l|l|}
\hline
  Transverse coupling: $\Phi _{x} = 0$ &Longitudinal coupling: $\Phi _{x} = \frac{k \pi}{2}$  \\ \hline
  $\eta^T = \frac{E_{J1} + E_{J2} }{2k}  \left( \frac{2\pi}{\Phi_0} \right)^2 L $ & $\eta^L = 0$ \\ \hline
   $E^{*T}_{Jq} = E_{Jq} + \left( \frac{\Phi_0}{2\pi} \right)^2 \frac{1+\eta^T}{2L}$  & $E^{*L}_{Jq} = E_{Jq} + \left( \frac{\Phi_0}{2\pi} \right)^2 \frac{1}{2L}$   \\ \hline
   $\omega_r^T= \sqrt{\frac{1 + \eta^T }{LC}}$ &  $\omega_r^L= \sqrt{\frac{1 }{LC}}$ \\ \hline
  $\omega_0^T = \sqrt{8 E_c E_{Jq}^{*T}} - E_c \frac{E_{Jq} +\left( \frac{\Phi_0}{2\pi} \right)^2 \frac{\eta^T}{2k^2L} }{E_{Jq}^{*T}} $ &  $\omega_0^L = \sqrt{8 E_c E_{Jq}^{*L}} - E_c \frac{E_{Jq} }{E_{Jq}^{*L}} $\\ \hline
  $g_{xx}^T =\frac{E_{J1} - E_{J2} }{2k^2}   \sqrt[4]{\frac{2 E_C}{E_{Jq}^{*T} }}  \frac{\pi}{\Phi_0} \sqrt[4]{\frac{L}{C}\frac{1}{1 + \eta^T}} $ & $g_{xx}^L =0$ \\ \hline
   $g_{zz}^T =-\frac{E_{J1} - E_{J2}}{16\,k^3}   \sqrt{\frac{2 E_C}{ E_{Jq}^{*T}}}  \left(\frac{\pi}{\Phi_0}\right)^2 \sqrt{\frac{L}{C}\frac{1}{1 + \eta^T}} $ & $g_{zz}^L = 0$ \\ \hline
   $g_{zx}^T =0$  & $g_{zx}^L =-\frac{E_{J1} - E_{J2}}{8\,k^2 }   \sqrt{\frac{2 E_C}{E_{Jq}^{*L}}}  \frac{\pi}{\Phi_0} \sqrt[4]{\frac{L}{C}} $   \\ \hline
   $g_{xz}^T =0$ &  $g_{xz}^L =-\frac{E_{J1} - E_{J2}}{4\,k^2}   \sqrt[4]{\frac{2 E_C}{ E_{Jq}^{*L}}}  \left(\frac{\pi}{\Phi_0}\right)^2 \sqrt{\frac{L}{C}} $ \\ \hline

\end{tabular}
\label{tab2}
\end{table}

We consider two forms of the magnetic flux modulation: a square-wave  and a sinusoidal one. Let us first focus on the case of a square-wave modulation of the magnetic flux


\begin{equation}
\begin{split}
\label{flux_mod}
\Phi _{x}(t) = \frac{k \pi}{2} \theta \left(  \cos \left( \varpi_s t  + \frac{3\pi}{2} \right) \right) ,
\end{split}
\end{equation}


\noindent
where $\theta \left( \cdot \right)$ is the Heaviside function which switches on periodically with period $T_s=1/\varpi_s$, where $\varpi_s$ is the frequency of the switching of the magnetic flux. By switching the external magnetic flux $\Phi _{x}(t)$ between the values $0$ and $\frac{k \pi}{2}$, one can tune the qubit and the resonator parameters in Hamiltonian (\ref{full_H}) at each instant of time. This gives the instantaneous switching between transverse and longitudinal qubit/resonator coupling which can be used to give rise to the dynamical Lamb effect.

In particular, for $\Phi _{x} = 0$ we can write the Hamiltonian (\ref{full_H}) {as}


\begin{equation}
\begin{split}
\label{H_trans_circ}
\hat{\mathcal{H}}_T = \hbar \omega_r^T \left( \hat{a}^{\dagger}\hat{a} + \frac{1}{2}\right) +
\hbar \frac{\omega_0^T}{2} \hat{\sigma}_z + \hbar g_{xx}^T \hat{\sigma}_{x} \left( \hat{a}%
^{\dagger} + \hat{a} \right) + \hbar g_{zz}^T \hat{\sigma}_{z} \left( \hat{a}%
^{\dagger} + \hat{a} \right)^2 ,
\end{split}%
\end{equation}


\noindent
where the expression of the parameters $\left\{  \omega_r^T, \omega_0^T, g_{xx}^T,  g_{zz}^T \right\}$ are given in Table \ref{tab2}. In this case, $\left\{ g_{xx},g_{zz}\neq 0; g_{zx},g_{xz}=0\right\}$ and the Hamiltonian (\ref{H_trans_circ}) is instantaneously equivalent to the Hamiltonian (\ref{transverse}) of a transversely coupled qubit/resonator system, with the exception of an extra coupling term.

{On the other hand, for $\Phi _{x} = \frac{k \pi}{2}$, Hamiltonian (\ref{full_H}) {can be reduced to the following form}}


\begin{equation}
\begin{split}
\label{H_long_circ}
\hat{\mathcal{H}}_L = \hbar \omega_r^L \left( \hat{a}^{\dagger}\hat{a} + \frac{1}{2}\right) +
\hbar \frac{\omega_0^L}{2} \hat{\sigma}_z  + \hbar g_{zx}^L \hat{\sigma}_{z} \left( \hat{a}%
^{\dagger} + \hat{a} \right) + \hbar g_{xz}^L \hat{\sigma}_{x} \left( \hat{a}%
^{\dagger} + \hat{a} \right)^2 ,
\end{split}%
\end{equation}


\noindent
where the expressions of $\left\{  \omega_r^L, \omega_0^L, g_{xx}^L,  g_{zz}^L \right\}$ are also given in Table \ref{tab2}. Here, $\left\{ g_{xx},g_{zz}=0; g_{zx},g_{xz}\neq 0\right\}$, which leads to an instantaneous longitudinal qubit/resonator coupling as in (\ref{longitudinal}), with a spurious coupling term. To suppress the unwanted terms $g_{zz}^T$ and $g_{xz}^L$ in Hamiltonian (\ref{H_trans_circ}) and (\ref{H_long_circ}), respectively, we choose specific values of the parameters of the circuit.

\subsection{Sinusoidal modulation}
\label{sin}

Although the square-wave modulation of the magnetic flux $\Phi _{x}(t)$ comes closest to the requirement of periodic and instantaneous switching "on" and "off" of the qubit/resonator coupling needed to observe the dynamical Lamb effect, this may be unrealistic in the experimental setting. For this reason, we turn to another type of modulation, a sinusoidal one, which can be easily obtained in experiments. In fact, a high-frequency sinusoidal magnetic flux was used in the first experimental observation of the dynamical Casimir effect \cite{wilson}. This models the more realistic situation where a finite amount of time is needed to switch "on" and "off" the coupling of the qubit with the resonator. Thus, we take $\Phi _{x} (t)$ as


\begin{equation}
\begin{split}
\label{flux_mod_sin}
\Phi _{x}(t) = \frac{k \pi}{2} \left( \frac{1}{2} + \frac{1}{2} \cos \left( \varpi_s t  \right) \right) .
\end{split}
\end{equation}


\noindent
In this case, the magnetic flux doesn't instantaneously switch "on" and "off" but continuously increases or decreases to its maximum or minimum value, respectively. However, the rise time $t_{\text{rise}} = t\left( \Phi _{x} = \frac{k \pi}{2} \right) - t\left( \Phi _{x} = 0 \right)$, that is the time required to increase the magnetic flux from the minimum value to the maximum value, and, vice versa, the fall time $ t_{\text{fall}} = t\left( \Phi _{x} = 0 \right) - t\left( \Phi _{x} = \frac{k \pi}{2} \right)$, the time needed to decrease it from the maximum value to the minimum value, are shorter than any parameter with dimension of time $(t_{\text{rise}}, t_{\text{fall}} \ll \omega_0^{-1},  \omega_r^{-1} )$. Therefore, one can still consider this modulation to be nonadiabatic. The parameters of Hamiltonian (\ref{full_H}) do not take the simple form shown in Table \ref{tab2} for the case of square-wave modulation but vary continuously with the magnetic flux $\Phi_x(t)$. These parameters can be found by substituting the sinusoidal modulation of the magnetic flux in the corresponding expressions from Table \ref{tab1} in the Appendix.

\section{results and discussion}
\label{res}

\begin{figure}[b]
	\subfloat[]{
	\includegraphics[width=7.3cm]{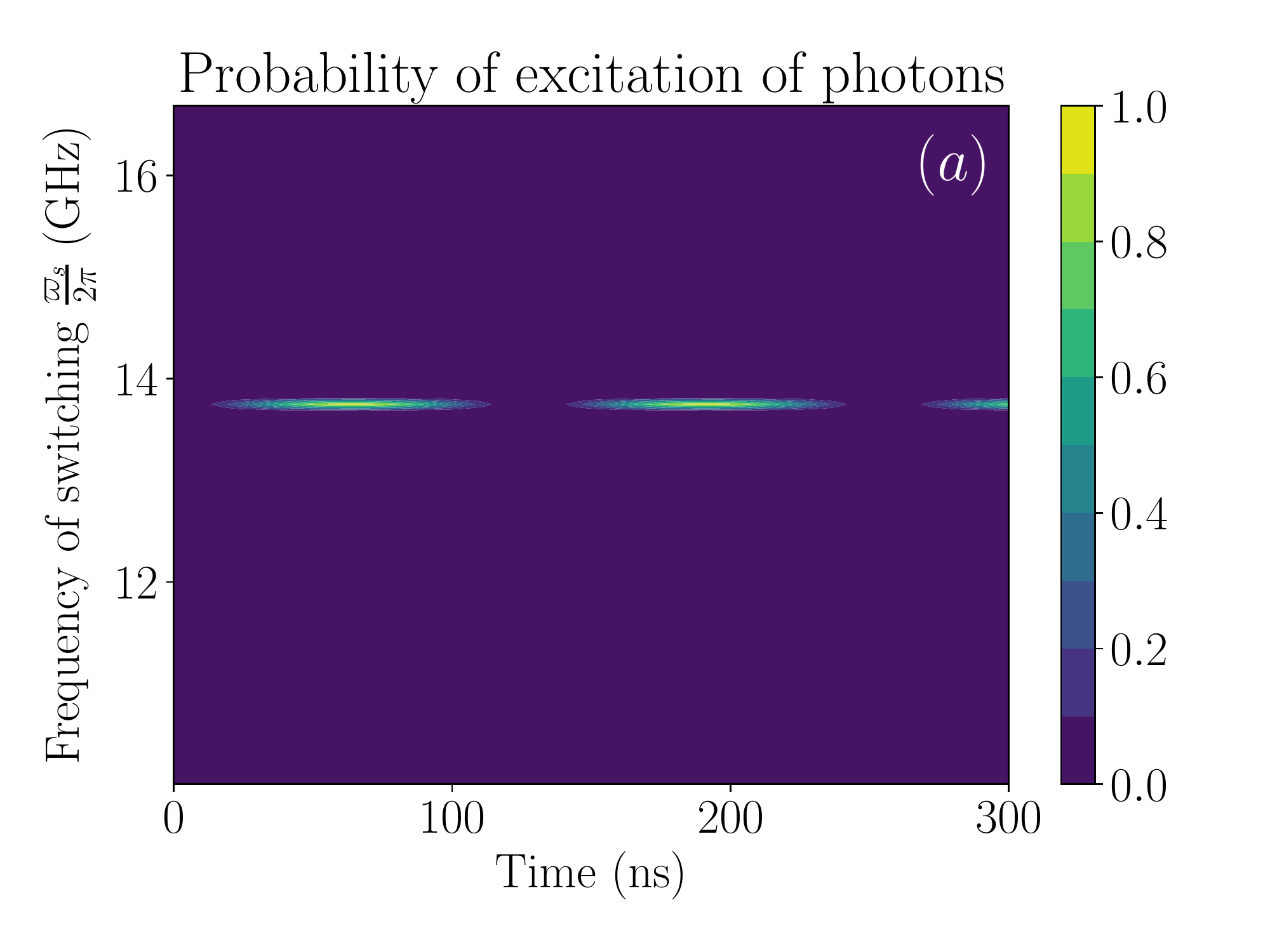}
	\label{fig:res}
	}
	\subfloat[]{
	\includegraphics[width=7.3cm]{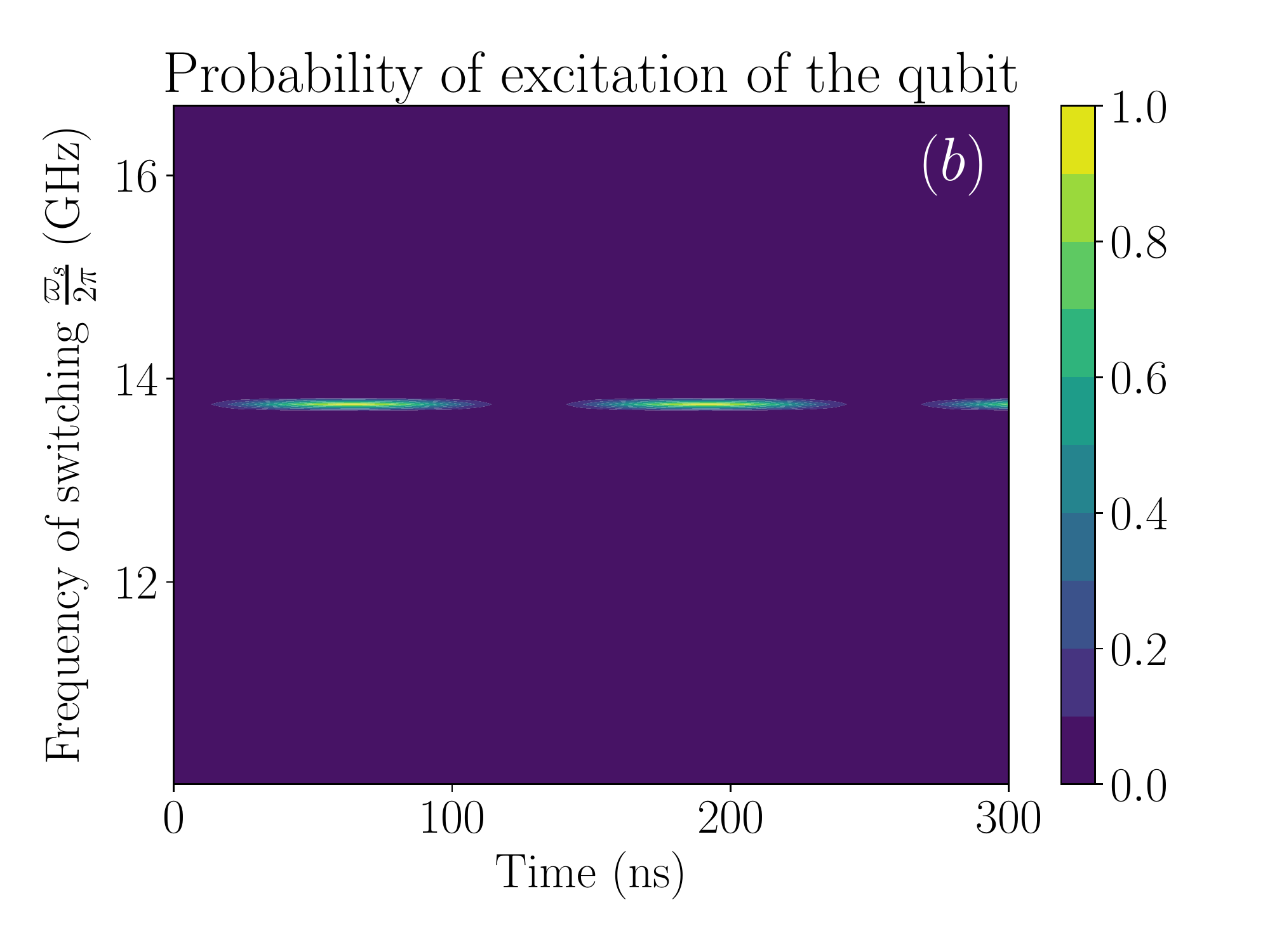}
	\label{fig:qub}
	} \\
	
	
	\subfloat[]{
	\includegraphics[width=7.3cm]{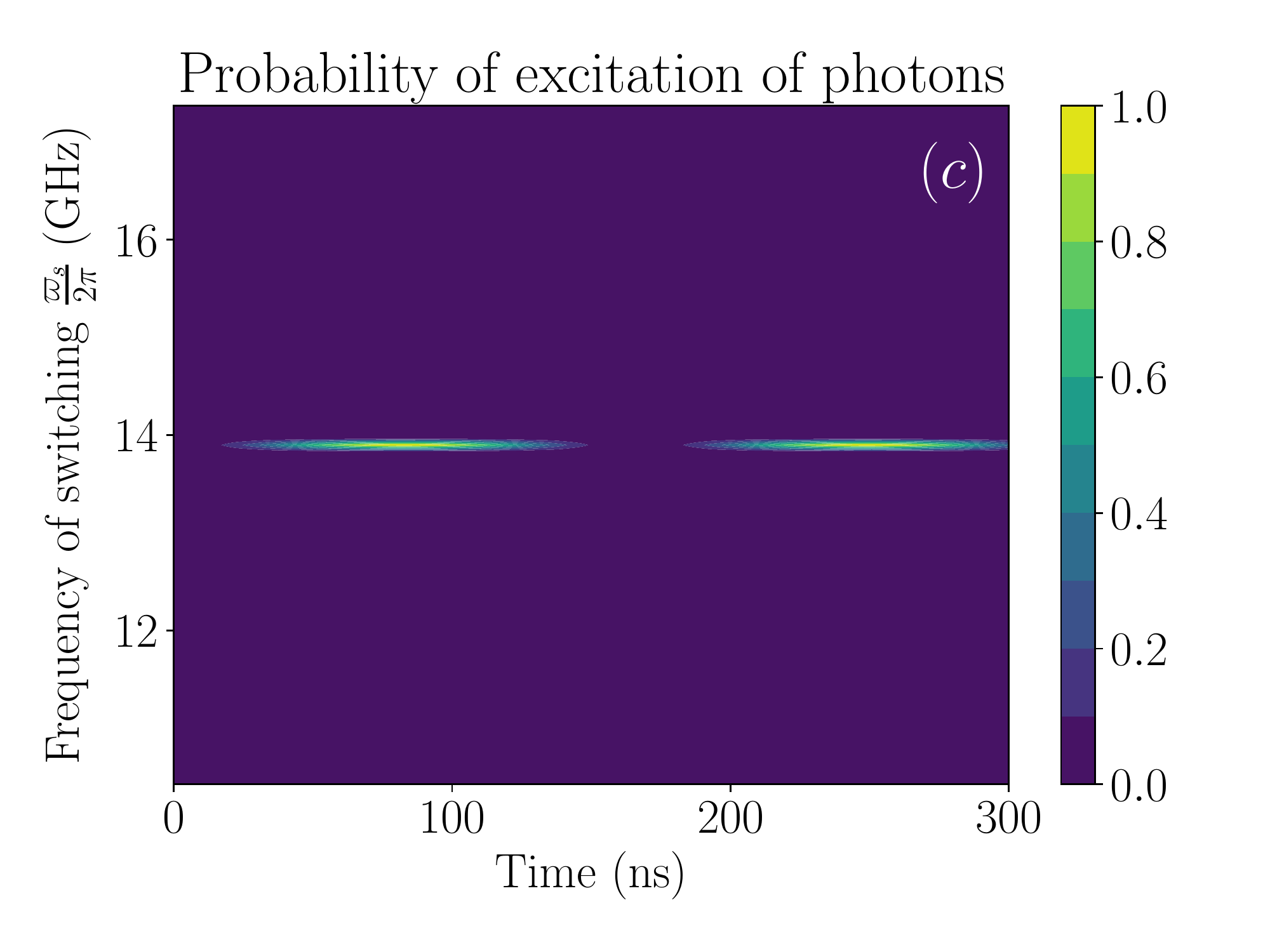}
	\label{fig:res2}
	}
	\subfloat[]{
	\includegraphics[width=7.3cm]{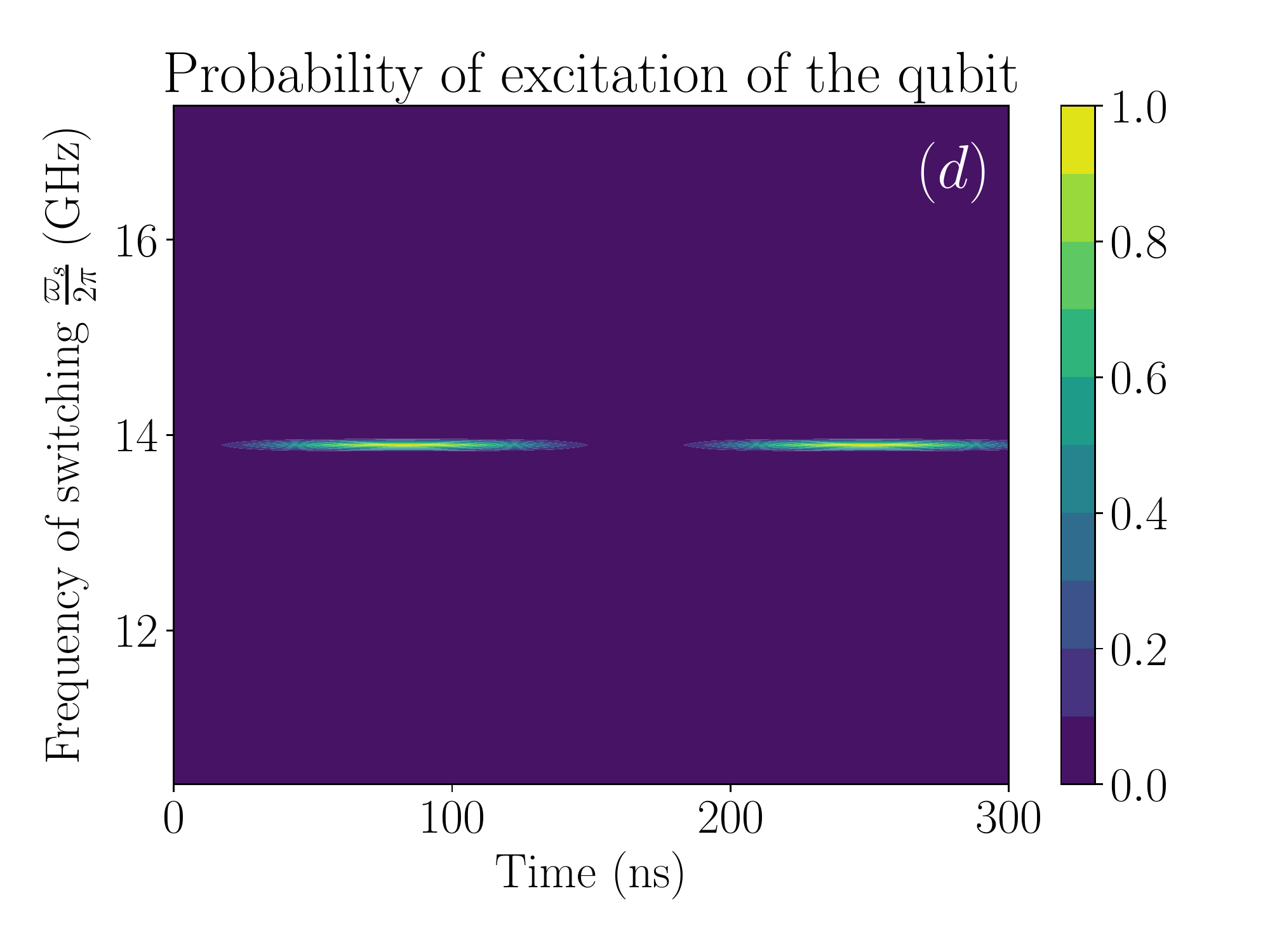}
	\label{fig:qub2}
	}
\caption{Time dependence of the probabilities of excitation of the qubit and the photons in the resonator for a range of frequencies of switching $\varpi_s$ of the magnetic flux. We take $\varpi_s \in \left[ \frac{3}{4} \left(\bar{\omega}_r+ \bar{\omega}_0 \right) , \frac{5}{4}\left( \bar{\omega}_r + \bar{\omega}_0 \right)  \right]$. The color-scale in the figures indicates the value of the probability. Probability of excitation of photons in the resonator (a) and the qubit (b), for a square-wave modulation of the magnetic flux. Probability of excitation of photons (c) and the qubit (d), for a sinusoidal modulation of the magnetic flux.
}
\label{fig:result}
\end{figure}

We numerically solve the Schr\"{o}dinger equation for the Hamiltonian (\ref{full_H}) in the case of periodic
switching between transverse and longitudinal coupling with the initial condition $\lvert \psi (t=0) \rangle = \lvert g, 0
\rangle $, where $g$ denotes the qubit in the ground state and $0$ is the number of photons in the resonator. In the numerical calculations of the probabilities of excitation of the qubit and creation of photons, we use the following values of the parameters of the circuit \cite{richer}: 
 
\begin{center}
\begin{tabular}{lcl}
$k=9,$ & $\; \; \; \; \;$ &$E_{Jq}=h\times 10$ GHz, \\
$E_{J 1} =  h\times 81.6$ GHz, & &$E_{J2}  =h\times  78.4$ GHz, \\
$C= 102$ fF, & &$C_q = 60$ fF,  \\
 $L=5$ nH. & &
\end{tabular}
\end{center}
 
Fig. \ref{fig:result} shows the time dependence of the probabilities of excitation of the qubit and the photons for a range of  frequencies of switching $\varpi_s$ of the magnetic flux. Clearly, there is a particular value of the switching frequency which corresponds to the maximal probabilities of excitation of the qubit and the photons. Figs. \ref{fig:res} and \ref{fig:qub} depict the results obtained in the case of a square-wave modulation of the magnetic flux, while the results obtained in the case of sinusoidal modulation of the magnetic flux are shown in Figs. \ref{fig:res2} and \ref{fig:qub2}.
In both cases, the value of the frequency of switching of the magnetic flux which maximize the probability of excitation of the qubit and the photons is $\varpi_s =  \bar{\omega}_r + \bar{\omega}_0$, which is the sum of the time-averaged qubit transition frequency $\bar{\omega}_0 = \frac{1}{T} \int_0^{T} \omega_0 \left( t' \right) dt' $ and the time-averaged photon transition frequency $\bar{\omega}_r =\frac{1}{T} \int_0^{T} \omega_r \left( t' \right) dt'$ over a period of oscillation of the magnetic flux. Because of the different time-dependence of the qubit and resonator transition frequencies for the different modulations, the probability of excitation of the qubit and the photons reach their maximum value at a different frequency of switching of the magnetic flux. In the case of a square-wave modulation, the probabilities are maximum for $\varpi_s =  \bar{\omega}_r + \bar{\omega}_0 = 13.75$ GHz. While for the case of a sinusoidal modulation, the maximum is at $\varpi_s = 13.90$ GHz. Moreover, the probabilities are constant with time and close to zero for almost all other values of the frequency of switching of the magnetic flux different from $\varpi_s = \bar{\omega}_r + \bar{\omega}_0$.

It is crucial to note that the state $\vert e,1 \rangle$, where $e$ stands for the qubit in the excited state, can only be reached from the initial state $\vert g,0 \rangle$ through the counter-rotating terms $\hat{a}^{\dagger}\hat{\sigma}^{+} + \hat{a}\hat{\sigma}^{-}$ in {Eq.} (\ref{full_H}). Since the counter-rotating terms are responsible for the presence of the Lamb shift of the qubit's energy level, the sudden change in Lamb shift can be obtained by nonadiabatically switching these terms "on" and "off". Also, when the qubit/resonator coupling is periodically switched "on" and "off" at a frequency equal to the sum of the qubit and the resonator average frequencies, the contribution of counter-rotating terms becomes important. This makes the dynamical Lamb effect the main channel of excitation of the qubit and the creation of photons. Thus, the $\vert g, 0 \rangle \rightarrow \vert e,1 \rangle$ transition caused by the nonadiabatic change of qubit/resonator coupling demonstrates the presence of the dynamical Lamb effect. A comparison of Figs. \ref{fig:res} and \ref{fig:qub}, and Figs. \ref{fig:res2} and \ref{fig:qub2}, clearly shows that the probabilities of excitation of the photon and the qubit coincide, indicating that the system is undergoing such transition.

\section{conclusion}
\label{conc}

{In conclusion, we predict that the dynamical Lamb effect could arise in superconducting circuits when the coupling of a superconducting qubit with a resonator is periodically switched "on" and "off" nonadiabatically and demonstrate that by using a superconducting circuit which allows to switch between longitudinal and transverse coupling of a qubit to a resonator, it is possible of to observe the dynamical
Lamb effect. In particular, the switching between longitudinal and transverse coupling which gives rise to the dynamical Lamb effect is achieved by turning "on" and "off" the magnetic flux through the loops of the superconducting circuit. If the magnetic flux is periodically turned "on" and "off" as a square-wave or a sinusoidal modulation with a frequency of switching equal to the sum of the average qubit and photon transition frequencies, the calculated probabilities of excitation of the qubit and the photons due to the dynamical Lamb effect reach their maximum values. }



\vspace{-0.4cm}
\acknowledgments
The authors are grateful to M. Kumph, D. C. Mckay and L. Glazman for the valuable and stimulating discussions.

\vspace{-0.4cm}
\appendix

\section{}

\label{appA}

The analytical expressions of the parameters for the Hamiltonian (\ref{full_H}) used in the calculations of the time-evolution of the probability of excitation of the qubit and photons are given in the table below \cite{richer}.

\begin{table}[h]

\caption{Expressions of the parameters introduced in Eq. (5). }
\centering
\begin{tabular}{|l|l|l|}
\hline
  $\eta(t) = \frac{E_{J1} + E_{J2} }{2k}  \left( \frac{2\pi}{\Phi_0} \right)^2 L \cos \left( \frac{\Phi_x(t)}{k} \right)$& $g_{xx} (t) =\frac{E_{J1} - E_{J2} }{2k^2}   \sqrt[4]{\frac{2 E_C}{E_{Jq}^*(t)}}  \frac{\pi}{\Phi_0} \sqrt[4]{\frac{L}{C}\frac{1}{1 + \eta(t)}} \cos\left(\frac{\Phi_x (t)}{k}\right)$ \\ \hline
   $E_c = \frac{e^2}{2C_q + C}$ & $g_{zz}(t) =-\frac{E_{J1} - E_{J2}}{16\,k^3}   \sqrt{\frac{2 E_C}{ E_{Jq}^*(t)}}  \left(\frac{\pi}{\Phi_0}\right)^2 \sqrt{\frac{L}{C}\frac{1}{1 + \eta(t)}} \cos\left(\frac{\Phi_x (t)}{k}\right)$ \\ \hline
   $E^{*}_{Jq}(t) = E_{Jq} + \left( \frac{\Phi_0}{2\pi} \right)^2 \frac{1+\eta(t)}{2L}$  & $g_{zx}(t) =-\frac{E_{J1} - E_{J2}}{8\,k^2 }   \sqrt{\frac{2 E_C}{E_{Jq}^*(t)}}  \frac{\pi}{\Phi_0} \sqrt[4]{\frac{L}{C}\frac{1}{1 + \eta(t)}} \sin\left(\frac{\Phi_x (t)}{k}\right)$   \\ \hline
   $\omega_r (t)= \sqrt{\frac{1 + \eta (t) }{LC}}$ &  $g_{xz}(t) =-\frac{E_{J1} - E_{J2}}{4\,k^2}   \sqrt[4]{\frac{2 E_C}{ E_{Jq}^*(t)}}  \left(\frac{\pi}{\Phi_0}\right)^2 \sqrt{\frac{L}{C}\frac{1}{1 + \eta(t)}} \sin\left(\frac{\Phi_x (t)}{k}\right)$ \\ \hline
  $\omega_0(t) = \sqrt{8 E_c E_{Jq}^*(t)} - E_c \frac{E_{Jq} +\left( \frac{\Phi_0}{2\pi} \right)^2 \frac{\eta(t)}{2k^2L} }{E_{Jq}^*(t)} $ &  \\ \hline
\end{tabular}
\label{tab1}
\end{table}

\

\end{document}